\documentclass[12pt,preprint]{aastex}

\shorttitle{NSVS01031772}
\shortauthors{L\'opez-Morales et al.}

\begin{document}

\title{NSVS01031772: A New 0.50+0.54 $M_{\sun}$ Detached Eclipsing Binary}

\author{Mercedes L\'opez-Morales\altaffilmark{1,7}, 
Jerome A. Orosz\altaffilmark{2}, 
J. Scott Shaw\altaffilmark{3,4}, 
Lauren Havelka\altaffilmark{2}, 
Maria Jes\'us Ar\'evalo\altaffilmark{5}, 
Travis McIntyre\altaffilmark{6}, Carlos L\'azaro\altaffilmark{5}}

\email{mercedes@dtm.ciw.edu, orosz@sciences.sdsu.edu, jss@hal.physast.uga.edu, havelka@sciences.sdsu.edu, mam@iac.es, travis@clemson.edu, clh@iac.es}

\altaffiltext{1}{Carnegie Fellow. 
Carnegie Institution of Washington, Department of Terrestrial
Magnetism, 5241 Broad Branch Rd. NW, Washington D.C., 20015, USA}
\altaffiltext{2}{Department of Astronomy, 
San Diego State University, San Diego, CA, 92182, USA}
\altaffiltext{3}{Department of Physics and Astronomy, 
University of Georgia, Athens, GA, 30602, USA}
\altaffiltext{4}{Southeastern Association for Research in Astronomy}
\altaffiltext{5}{Dept. de Astrof\'isica, Universidad de La Laguna; 
Instituto de Astrof\'isica de Canarias, E38205, La Laguna, Tenerife, Spain}
\altaffiltext{6}{Department of Physics and 
Astronomy, Clemson University, Clemson, 
SC, 29634, USA}
\altaffiltext{7}{Visiting Astronomer, 
Kitt Peak National Observatory, NOAO, which is operated by 
the Association of Universities for Research in Astronomy, Inc. 
(AURA) under cooperative agreement with the National Science Foundation.}

\begin{abstract}
We report the discovery of a new detached eclipsing binary 
known as NSVS01031772 with component masses
$M_1$= 0.5428 $\pm$ 0.0027$M_{\sun}$, $M_2$= 0.4982 $\pm$
0.0025$M_{\sun}$, and radii $R_1$= 0.5260 $\pm$ 0.0028$R_{\sun}$,
$R_2$= 0.5088 $\pm$ 0.0030$R_{\sun}$. The system has an orbital period
of 0.3681414(3)  days and an apparent magnitude V $\simeq$
12.6. The estimated effective temperatures of the stars are $T_{{\rm
eff}_1} = 3615 \pm 72K$ and $T_{{\rm eff}_2} = 3513 \pm 31K$. The space velocities of
the system suggest that it is a main sequence binary and with a
metallicity that is approximately  solar. The two stars in this binary
are located in a region of the Mass-Radius relation where no accurate
observational data was previously available. Similarly to other
low-mass binaries recently studied, the radius of each star in
NSVS01031772 exceeds the best evolutionary model predictions by about 8.5\% on average.
\end{abstract}


\section{Introduction} \label{sec:intro}

Low-mass stellar structure models have improved fast in the past few
years as reviewed by Chabrier \& Baraffe (2000) and Chabrier et
al.~(2005). However, that improvement has not yet been matched by
observations. Accurate parameters of low-mass stars are difficult to
obtain, with the best source of precise data being double-lined
eclipsing binaries (DDEBs), but those systems are scarce. Only two
low-mass DDEBs had been found until 1998, CM Dra (Lacy 1977; Metcalfe
et al.~1996) and YY Gem (Leung \& Schneider 1978; Torres \& Ribas
2002). A few more have been discovered since then, CU Cnc (Delfosse et
al.~1999; Ribas 2003), BW3 V38 (Maceroni \& Montalb\'an 2004),
TrES-HerO-07621 (Creevey et al.~2005), GU Boo (L\'opez-Morales \&
Ribas 2005), 
2MASS J05162281+2607387 (Bayless \& Orosz 2006), 
UNSW-TR-2 (Young et al. 2006),
and a new binary found in the open cluster NGC1647 by
Hebb et al.~(2006). However, the number of low-mass DDEBs is still small,
and the error bars on the parameters of some of those systems are too large 
to place rigorous constraints on the evolutionary models. There are also
portions of the Mass-Radius relation where no good observations exist
(e.g.~0.25--0.35 and 0.5--0.6$M_{\sun}$).

Clearly, the sample size of well-studied low-mass binaries
needs to be increased.
Over the past two years we have searched for candidate low-mass binaries 
in the Northern Sky Variability Survey (NSVS)
database (Wozniak et al. 2004). Our search algorithm uses two period-searching 
methods: the ``string/rope'' method based on the Lafler-Kinman 
statistic (Clarke 2002), and
the analysis of variance method (Schwarzenberg-Czerny 1989). 
The first candidate system,  known as NSVS01031772 (hereafter NSVS0103),
with J2000 coordinates of  $\alpha=$~13:45:35, $\delta=+79$:23:48 (Figure 1),
was identified as a possible low-mass binary 
based on its short orbital period
($P\approx0.368$ days), the duration of its eclipses 
($\approx 16$\% of phase), and its red
near-infrared colors, obtained from the 2 Micron All Sky Survey (2MASS)
database
(Skrutskie et al. 2006).
In this work we present follow-up spectroscopic and
photometric observations which confirm the low-mass nature of NSVS0103.
We derive accurate fundamental parameters for the component stars
and compare our results to evolutionary models.

\section{Radial Velocity Curves} \label{sec:rv}

We collected a total of 108 spectra 
in two nights during May 2005 with the
echelle spectrograph at the 4-m Mayall telescope at the Kitt Peak
National Observatory (KPNO). The wavelength coverage of each spectrum
is 5700--8160~\AA, with a resolving power of 18,750 at 6000~\AA\ and
an average signal-to-noise ratio (S/N) of 7--10
per pixel. We also obtained a high S/N spectrum of the
M dwarf GJ 740 (M1.5V) for use as a template in the derivation of the
radial velocities. By using a real star as template we avoid the
problems that the low-mass stellar atmosphere models have reproducing
some spectral features of the stars.

The radial velocities were extracted using the implementation of the
cross-correlation algorithm TODCOR (Zucker \& Mazeh 1994) kindly
supplied by Guillermo Torres. The analysis of NSVS01031 with TODCOR
was analogous to those described by Torres \& Ribas (2002) and
L\'opez-Morales \& Ribas (2005) in their analyses of YY Gem and GU Boo.
The template was rotationally broadened by 70 km~s$^{-1}$ to
optimize the results of TODCOR. The standard errors of the
resulting radial velocities are 10 km~s$^{-1}$ or less. We tested
for systematic effects in the derived velocities
and found no obvious ones.
The radial velocities for both stars phased
on the eclipse ephemeris derived below  are shown in Figure~\ref{fig:rv}.
The data on that figure are available online. A sample of the contents of the 
online table is shown in Table 1.

\section{Light Curves}\label{sec:lc}

We obtained complete $V$, $R$, and $I$-band light curves of NSVS0103
over 12 nights between March and May 2005, using the Apogee U55
512x1152 CCD on the Southeastern Association for Research in Astronomy
(SARA) 0.9-m telescope at KPNO. The data were 
reduced and analyzed using standard
aperture photometry packages in IRAF\footnote{IRAF is distributed by
the NOAO, which are operated by the Association of Universities for
Research in Astronomy, Inc., under cooperative agreement with the
NSF.}, with no differential extinction effects taken into account
given the relative small separation between the target and the
comparison and check stars in the field
(see Figure 1). The final light curves, which
contain 649 points in $V$, 843 points in $R$, and 1210 points
in $I$, are illustrated in Figure 3. The average photometric
precision per data point is 0.017 mags in $V$, 0.013 mags in $R$, and 0.007 mags in $I$.

NSVS0103 was also observed on the nights of June 29 and 30, 2006
using the
0.6m telescope at the Mount Laguna Observatory (MLO), equipped with an SBIG
CCD and $R$ and $I$ filters.  IRAF was again used to apply the flat-fielding and
dark current corrections, and to derive differential light curves.
The MLO light curves, which contain
366 points in $R$ and 364 points in $I$, are shown in Figure 4.  The average photometric
precision per data point is in this case 0.009 mags in $R$ and 0.0011 mags in $I$. All the data shown in figures 3 and 4 are available online. Table 2 shows a sample of the contents of the online table.

Times of minimum light derived from our photometry and from the NSVS
database are given in Table 3.

From those times of minima we derive the following ephemeris equation
\begin{equation}
T_{Min I} = \mbox{HJD}\, 2,453,456.6796(2) + 0.3681414(3)E
\end{equation}
With the exception of one point, the typical ($O-C$) residuals are 
less than one minute.  There are no obvious trends.
Using the orbital
period above, we find an average phase difference between primary and
secondary minima of $\Delta\phi$ = 0.4989$\pm$0.0027,
which is consistent with a circular orbit.

\section{Analysis} \label{sec:anal}

Each set of light curves was simultaneously modeled with the velocity curves
using the ELC code (Orosz \& Hauschildt 2000) and its various optimizers 
using updated model atmospheres for
low-mass stars and brown dwarfs (Hauschildt, priv.~comm.).
The model has two main types of free parameters: those
related to the geometry of the binary such as the stellar masses,
radii, and separation, and those related to the radiative
properties of the stars such as their effective temperatures,
gravity and limb darkening, and locations of spots (if any).  
For the geometrical parameters,  we assumed a fixed orbital period
of 0.3681414 days, zero eccentricity, and synchronous rotation for
both stars. The free geometrical parameters
were the inclination $i$, the mass and radius of the primary,
$M_1$ and $R_1$, the ratio of the radii $R_2/R_1$,
and the $K$-velocity of the primary $K_1$.  For a given inclination
$i$ and orbital period $P$, the orbital separation $a$ and
the mass ratio $Q=M_2/M_1$ can be found specifying  $M_1$ and $K_1$.  
The value of $a$ sets the absolute scale of the binary, while $Q$ 
gives a unique Roche geometry. We find that $M_1$ and $K_1$ are more 
efficient to optimize 
the solutions for well-detached binaries than $a$ and $Q$,
since $K_1$ can be inferred from the radial velocity
curve and $M_1$, roughly, from the spectral type. Once the scale 
and the mass ratio are known, specifying $R_1$ and the ratio of the 
radii
$R_2/R_1$ gives the specific values of the equipotentials.  

For the radiative properties of the stars, we eliminate the need for a
parameterized limb darkening law by using tabulated  model atmosphere with
specific intensities over a wide range of temperatures and gravities.  
The gravity darkening
exponents were set according to the temperatures of the stars following
Claret (2001).  The stars in NSVS0103 are nearly spherical
so there is essentially no gravity darkening. We used 
``simple reflection'' (see Wilson 1990).  The free radiative-properties 
parameters were the temperature of
the primary $T_1$, the temperature ratio $T_2/T_1$, and
the parameters to describe two spots on the primary (this is the model that provided the best results in both the SARA and MLO datasets, see below).
The spots in ELC are specified by a temperature factor, the 
longitude and latitude of the spot center, 
and the angular radius of the spot, similarly to the Wilson-Devinney 
code (1971). Finally, ELC has a phase-shift parameter to account for
small errors in the ephemeris. In total, we have 16 free parameters.

We began by modelling the radial velocity curves
simultaneously with the SARA light curves since these light
curves have better sampling and more filters than the MLO light curves.
It became immediately clear that the model radial velocity curves did not
match the observed radial velocities near the conjunction
phases (e.g.\ during the primary and secondary eclipses).
During partial eclipse, the spectral line profiles may no longer
symmetric, which may lead to a difference between the measured
radial velocity and the actual radial velocity of the star.
ELC computes the change in radial velocity
(generally known as the Rossiter effect) using the technique outlined
in Wilson \& Sofia (1976), which accounts for the shift in the
``center of light''.  Since this simple technique may not fully mimic
the way TODCOR measures velocities during partial eclipse, we
did not include in the fit the primary and secondary radial velocities 
between phases 0.425--0.575  and 0.925--1.075.  In all, a total of 38 points
were excluded from each radial velocity curve, leaving 71 points per curve.  
We note that since the mass constraints come mainly from observations
near the quadrature phases, excluding the points near the conjunction phases
will have very little if any effect on our results. The points not included 
in the fit are represented as open symbols in figure 2. 

The SARA light curves are not symmetric about phase 0.5. 
Usually asymmetries in the light curves are attributed to the presence
of spots on one or both stars.
We tried models with a single spot on
the primary and models with a single
spot on the secondary, but could not find an acceptable solution. 
We then tried models with two spots on the primary, models
with two spots on the secondary,
and models with a spot on each star.
The best 
solution was found with a model with two bright spots on the primary. That model is illustrated in the top diagram of Figure 5.
We run the ELC genetic optimizer 
for 700 iterations, using the final solution to scale the error bars so 
that $\chi^2=N-1$ for each data set. 
The resulting mean errors per point were 0.017 mag in $V$, 0.013 mag in $R$,
and 0.006 mag in $I$.  
After scaling the error bars, we ran the genetic 
code and also a simple $grid$ $search$ optimizer to establish the $1\sigma$ 
errors on the fitted  and derived parameters
using brute-force (see Orosz et al. 2002).  
The results are summarized in the second column of Table 4. 
Figures \ref{fig:rv} and \ref{fig:lcsSARA} show the best model
radial velocity and light curves plotted with the observations.

The MLO light curves were modeled in the same way as the SARA light curves
described above. The results are summarized in the third column of Table 4. 
Figure 4 shows the best model fits for those observations. 
The two bottom diagrams in figure 5 show the best spot configuration in this 
case.
The light curves from MLO turned out to be not as useful as the SARA
light curves for deriving precise parameters of the system.   
The statistical errors on the fitted and derived
parameters from the MLO data are typically a factor of three larger
than the statistical errors from the SARA data. 
The SARA light curves have a total of 2702
points in three filters obtained with a 0.9m telescope, whereas
the MLO light curves have a total of 730 points in two filters
obtained with a 0.6m telescope, so it is not surprising that 
much more precise results were obtained from the SARA data.

\section{Parameters of NSVS01031772}

We use the ELC solutions discussed
in \S 4 to derive the physical parameters
of NSVS0103. The masses of the stars derived from the orbital 
solution of the SARA data are $M_1$= 0.5416 $\pm$ 0.0068$M_{\sun}$ and 
$M_2$= 0.4988 $\pm$ 0.0048$M_{\sun}$. For the radii we obtain 
$R_1$= 0.5273 $\pm$ 0.0029$R_{\sun}$ and  $R_2$= 0.5058 $\pm$ 0.0032$R_{\sun}$.
The values of those parameters from the MLO data are 
$M_1$= 0.543 $\pm$ 0.003$M_{\sun}$ and  $M_2$= 0.498 $\pm$ 0.003$M_{\sun}$ 
and  
$R_1$= 0.510 $\pm$ 0.010$R_{\sun}$ and  $R_2$= 0.538 $\pm$ 0.010$R_{\sun}$.
Given the discrepancies between the two datasets, mainly in the values of the stellar radii, we decided to adopt a weighted average of those parameters as our final values. The results are shown in Table 5.

The projected rotational
velocities of the stars, derived from their radii and the orbital
period of the system (assuming synchronous rotation), are ${v_{\rm
sync}}_1 \sin i= 72.13\pm0.38$ and 
${v_{\rm sync}}_2 \sin i=69.77\pm0.41$ km s$^{-1}$, 
consistent with the results from
TODCOR. The values of $T_{eff1}$ and $T_{eff2}$ in Table 5 correspond to
the average of the SARA and MLO results. Note that the effective temperatures 
reported in Table 4 correspond to the temperature of the photosphere of the stars without spots. We recomputed the values of $T_{eff1}$ to account for the presence of spots. The resultant mean effective temperature of the system is therefore $T_{eff}= 3564 \pm 73K$. This value agrees within the errors with the mean effective temperature of $T_{eff}= 3750 \pm 185$K derived using the color-$T_{eff}$ relations in Table 2 of L\'opez-Morales \& Ribas (2005).

The luminosity and absolute  magnitude of the stars were
computed from their effective temperatures, radii, and surface
gravities using filter-integrated {\sc NextGen}
model atmospheres provided by France
Allard.  A simple computer code was written that uses
the temperature, gravity, and radius plus
the $1 \sigma$ errors of each component as
inputs to compute the absolute magnitudes of the binary 
in the standard UBVRIJHK filters.  We find,
for example, $M_V = 9.08\pm 0.22$,
$M_J = 5.77\pm 0.09$, and 
$M_K = 4.89\pm 0.08$.
The distance can be computed by using the observed near infrared
magnitudes given in the 2MASS database, where NSVS0103 has the designation
2MASS J13453489+7923482.   
The observed magnitudes are listed as $J = 9.692
\pm 0.021$, 
$H = 9.021 \pm 0.018$, 
and 
$K = 8.778 \pm 0.016$.  The $K$-band
extinction to infinity in this direction is 0.012 mag, according to the
extinction maps (see http://irsa.ipac.caltech.edu/applications/DUST/).
Assuming the temperatures above, and the best-fitting radii and gravities,
we find a distance of $d=60.3 \pm 1.0$ pc using the observed $K$ magnitude and
extinction and a distance of
$d=60.9 \pm 0.9$ using the $J$-band magnitude and
extinction. 

\subsection{Age, Space Velocities and 
Stellar Activity Indicators}\label{sec:age}

NSVS0103 does not seem to be related to any known cluster, stellar
association, or star formation region. Therefore, we can only evaluate
its age from its space velocities $(U,V,W)$\footnote{Positive values
of $U$, $V$, and $W$ correspond to velocities toward the Galactic
center, Galactic rotation and North Galactic pole.}. The heliocentric
space velocities of NSVS0103 were computed from its position, radial
velocity ($\gamma=19.0\pm1.0$ km~s$^{-1}$), distance ($d=60.6\pm 0.7$
pc), and proper motions; the latter retrieved from the USNO-B1.0
catalog (Monet et al.~2003): 
$\mu_{\alpha} =100 \pm 4$ and
$\mu_{\delta} =62 \pm 1$ mas yr$^{-1}$. The obtained components
of the space motion are
$U =-21.07 \pm 0.16$, 
$V =16.54 \pm 0.04$ and 
$W = -1.19  \pm 0.15$
km~s$^{-1}$, which correspond to a total space velocity of
$S =26.81 \pm 0.13$ km~s$^{-1}$. The value of $W$ indicates that NSVS0103
is confined to the galactic plane. The location of the binary in the
$U - V$ plane does not fall within the young disk stars area defined
by Eggen (1989), nor the area occupied by the young population tracers
(Skuljan et al.~1999). Also the space velocities of NSVS0103 do not
match any known moving group (Montes et al.~2001). We conclude therefore that
NSVS0103 is not a young object and has most likely already reached the
main sequence. In addition we expect the metal abundance of NSVS0103
to be close to solar, since its space motions agree with a disk
population.

NSVS0103 appears in the ROSAT All-Sky Bright Source Catalog (Voges et
al.~1999) as X-ray source 1RXS J134540.6+792332. Using the calibration
equation by Schmitt et al.~(1995) we estimate an X-ray 
flux of 
$(5.97 \pm 1.25) \times 10^{-13}$ ergs~cm$^{-2}$~$s^{-1}$,
which results in an X-ray
luminosity of
$\log L_{X} ({\rm ergs~s}^{-1}) = 29.39 \pm 0.09$, 
which is similar to the
X-ray luminosities of YY Gem, 
$\log L_{X} ({\rm ergs~s}^{-1}) = 29.27 \pm 0.02$, 
and GU Boo, 
$\log L_{X} ({\rm ergs~s}^{-1}) = 29.3 \pm 0.2$, 
computed using the same calibration equation. Finally,
we observe strong $H_{\alpha}$ emission lines in our spectra, with an
average equivalent width of $5.4\pm 1.1$~\AA\
over all phases. That
$H_{\alpha}$ emission level is higher than those of YY Gem 
(2.0~\AA), CU Cnc (3.85--4.05~\AA) and GU Boo (1.7~\AA).

\section{Comparison with Models} \label{sec:models}
We compare the masses and radii of the stars in 
NSVS0103 to the predictions by the models of 
Baraffe et al.~(1998) (hereafter B98) and 
Siess et al.~(2000) (hereafter S00), which are  
the only models that explicitly attempt to 
reproduce the properties of main sequence low-mass stars. We also include 
in this comparison the other known binaries compiled in Table 10 
of L\'opez-Morales \& Ribas (2005), and the new systems found by Hebb et al.~
(2006), Bayless \& Orosz (2006) and Young et al. (2006). After exploring a range of metallicities and ages, 
we conclude that the Z=0.02 models provide the closest 
values to the measured masses
and radii of NSVS0103. Different isochrones give  practically identical
results, assuming the system has already reached the main
sequence\footnote{Chabrier \& Baraffe (1995) estimate a time of
arrival to the ZAMS of about 0.3 Gyr for stars below 0.6 $M_{\odot}$},
so we chose a 0.35 Gyr isochrone which also allows us to compare
NSVS0103 to the other binaries. This comparison is illustrated in
Figure~\ref{fig:mr}. We find that the models consistently predict
smaller radii than the ones observed. The discrepancies are of the
order of 5\% for the primary and 11.9\% for the secondary for B98. In the case of the S00 models, those discrepancies are 15.5\% and 22\%. This result fully agrees with
the trend observed in previous studies of YY Gem, CU Cnc, and GU Boo. We also show in figure 6 the empirical Mass-Radius relation derived by Bayless \& Orosz (2006), using the parameters of previously studied low-mass binaries. The parameters of the secondary in NSVS0103 fit well in this case, but the radius of the primary is overestimated by about 5\%

We compare next the $T_{\rm eff}$ and $M_{V}$ of the stars in NSVS0103 to
the two models above. The error bars of these parameters are much
larger than in $M$ and $R$, since we have to use external calibrations
to compute their values. However, they are still useful to check for
significant discrepancies between models and observations. The
temperatures predicted by B98 for 0.50 and 0.54 $M_{\sun}$ stars are
3657K and 3750K, while the predictions by S00 for those same stars are
3810K and 3880K. In both cases the models predict hotter temperatures than the ones obtained for NSVS0103. Finally, B98 predict
absolute magnitudes of 9.89 and 9.48 for 0.50 and 0.54 $M_{\sun}$
stars, while the values predicted by S00 are 9.97 and 9.56. In both cases the absolute magnitudes of NSVS0103 agree with the models, given the uncertainties.

\section{Summary and Discussion} \label{sec:sum}

We have found a new detached eclipsing binary composed of two M-type
stars with masses between 0.49 and 0.53 $M_{\sun}$. Both stars fall in
a stretch of the mass-radius relation where no accurate data were
available before. The radii of these stars are the most accurate
derive to date for low-mass stars, with uncertainties of only 0.55 and
0.63\%. The estimated temperatures of the stars are 
$T_{{\rm eff}_1} =
3615 \pm 72K$ and $T_{{\rm eff}_2} = 3513 \pm 31K$, after accounting for the effect of
spots. We also conclude from the space velocities that the binary is
most likely on the main sequence stage. Therefore the parameters of
the stars are suitable to test stellar structure models.

The models that most closely reproduce the radius of the stars in
NSVS0103 are B98 ($Z=0.02$). However, they still underestimate the
observed values by 8.5\% on average; a result similar to previous 
studies of other
binaries. This discrepancy between models and observations could be
attributed to innacuracies in the equation of state. However, the same
models describe well stars in long period ($> 10$ days) spectroscopic
binaries and single field stars (Delfosse et al.~2000; Ribas
2005). One difference between short period binaries and the long
period binaries and single stars is the higher rotational velocities
of the former as result of orbital synchronization. The higher
rotational speeds enhance the magnetic activity, and hence the
appearance of large spots, X-ray activity, and emission lines in the
stars. The observed larger radii and lower temperatures could be a
consequence of that enhanced activity. Mullan \& MacDonald (2001), for
example, suggest that the larger radii may be caused by the inhibition
of convection in stars with strong magnetic fields. The stars would appear cooler as a consequence of the
larger radii. If this hypothesis is correct, it affects not only the
components of short period binaries, but also any active low-mass
star. If the radius of low-mass stars is correlated to their magnetic
activity (equivalently to their rotational velocity), the current
models may only apply to slow, inactive rotators. For younger, faster
rotating stars, their magnetic activity level may have to be included
in the models as an additional parameter to compute their fundamental
properties.

\acknowledgments

We thank the Southeastern Association for Research in Astronomy for
making their facilities available to us. We also thank G. Torres for
providing us with his implementation of TODCOR, and our referee for
useful comments and suggestions. M.~L-M. receives research and travel
support from the Carnegie Institution of Washington through a Carnegie
Fellowship. L.~H. was supported by NSF-REU grant AST-0453609 to
San Diego State University.
T.~M. was supported by NSF-REU grant AST-0097616 to the
Southeastern association for Research in Astronomy. This project was
partially supported by the National Aeronautics and Space
Administration grant NAG5-12182. We made use of data from the Northern
Sky Variability Survey created jointly by the Los Alamos National
Laboratory and University of Michigan. The NSVS was funded by the US
Department of Energy, the National Aeronautics and Space
Administration and the National Science Foundation.

\clearpage

\begin{table}[t]
\centering
\footnotesize
\caption{Sample of radial velocities table. A full version of this table is available online}
\label{tab:times}
\begin{tabular}{lrrrrr}
\hline\hline
HJD (days)  & phase &$K_{1}$ (km/s)&$K_{1}$ err (km/s)&$K_{2}$ (km/s) & $K_{2}$ err (km/s)\\
\hline
2453493.7683& 0.746 & 160.51 & 9.29 &-134.44 &14.74\\
2453493.7710& 0.753 & 158.48 &11.32 &-136.25 & 9.47\\
2453493.7737& 0.760 & 164.10 &14.24 &-135.40 &12.21\\
...         &  ...  & ...    & ...     & ...    & ... \\
\hline\hline
\end{tabular}
\end{table}

\begin{table}[t]
\centering
\footnotesize
\caption{Sample of SARA and MLO light curve data. A full version of this table is available online}
\label{tab:times}
\begin{tabular}{cccc}
\hline\hline
HJD (days)  & phase & mag    &mag err\\
\hline
2453456.92852& 0.676& 0.466& 0.018\\
2453456.92971& 0.679& 0.458& 0.016\\
2453456.93091& 0.683& 0.470& 0.016\\
2453456.93212& 0.686& 0.481& 0.016\\
2453456.93331& 0.689& 0.464& 0.016\\
...          &  ... & ...    & ... \\
\hline\hline
\end{tabular}
\end{table}

\begin{table}[t]
\centering
\footnotesize
\caption{Observed Times of Minimum Light}
\label{tab:times}
\begin{tabular}{ccc}
\hline\hline
Time of minima & cycle \#         & O-C \\
(HJD)          & (since $T_{0}$) & (min) \\
\hline
2453473.98266 & 47.0   &   $      0.596448$ \\
2453478.95221 & 60.5   &   $      0.079632$ \\
2453480.97698 & 66.0   &   $      0.068544$ \\
2453480.79303 & 65.5   &   $      0.242352$ \\
2453490.91543 & 93.0   &   $     -1.901088$ \\
2453492.94170 & 98.5   &   $      0.247824$ \\
2453493.86213 & 101.0  &   $      0.357984$ \\
2453520.73664 & 174.0  &   $      0.628416$ \\
2453520.92006 & 174.5  &   $     -0.308592$ \\
2453916.67214 & 1249.5 &   $     -0.200592$ \\
2453916.85669 & 1250.0 &   $      0.489600$ \\
2453917.77661 & 1252.5 &   $     -0.134640$ \\
\hline\hline
\end{tabular}
\end{table}

\begin{table}[t]
\centering
\footnotesize
\caption{Model Parameters for Light Curve Solution}
\label{tab:parm}
\begin{tabular}{lr@{\,$\pm$\,}lr@{\,$\pm$\,}l}
\hline\hline
Parameter & \multicolumn{2}{c}{Value from SARA} & 
            \multicolumn{2}{c}{Value from MLO}\\
\hline
orbital period $P$ (days) \dotfill & \multicolumn{4}{c}{fixed at 0.368141} \\
eccentricity $e$      \dotfill & \multicolumn{4}{c}{fixed at 0.0} \\
rotational velocities \dotfill & \multicolumn{4}{c}{fixed at synchronous} \\
gravity darkening, primary \dotfill & \multicolumn{4}{c}{0.0450244} \\
gravity darkening, secondary \dotfill & \multicolumn{4}{c}{0.0450994} \\ 
limb darkening coefficients \dotfill & \multicolumn{4}{c}{from model 
                                                        atmospheres} \\
reflection \dotfill & \multicolumn{4}{c}{``simple'' reflection} \\
inclination $i$ (deg)  \dotfill & 85.91&0.03  & 85.86&0.05 \\
mass ratio $Q\equiv M_2/M_1$ \dotfill & 0.9217&0.0048  &  0.9166&0.0050 \\
radius ratio $R_2/R_1$  \dotfill &  0.9638&0.0066   & 1.054&0.051 \\
temperature ratio $T_2/T_1$  \dotfill &  1.01129&0.00036  & 0.9918&0.0012  \\
primary temperature $T_1$ (K) \dotfill & 3505&6 & 3512&14 \\
secondary temperature $T_2$ (K) \dotfill & 3545&6 & 3482&18 \\
separation $a$ ($R_{\odot}$) \dotfill & 2.1870&0.0089  & 2.1910&0.0050  \\
$K_1$  (km s$^{-1}$) \dotfill & 143.85&0.37    & 143.5&0.6   \\
$K_2$  (km s$^{-1}$) \dotfill & 156.06&0.88    & 156.6&0.7   \\
Omega-potential 1 \dotfill & 5.121&0.015 & 5.24&0.11 \\
Omega-potential 2 \dotfill & 5.053&0.022 & 4.81&0.08  \\ 
temperature factor, spot 1 \dotfill & 1.048&0.004  & 1.19&0.02 \\
latitude spot 1 (deg)      \dotfill & 26.1&0.3  & 104&17\\
longitude spot 1 (deg) \dotfill    & 310.9&0.5  & 291.5&2.0 \\
angular radius spot 1 (deg) \dotfill & 88.4&0.5 & 8.5&1.0 \\
temperature factor, spot 2 \dotfill & 1.196&0.002 & 1.075&0.010 \\
latitude spot 2 (deg)      \dotfill & 42.4&1.0 & 46&6 \\
longitude spot 2 (deg) \dotfill    & 90.6&0.5 & 63.5&2.0\\
angular radius spot 2 (deg) \dotfill & 14.9&0.2 & 24.5&0.5 \\
\hline\hline
\end{tabular}
\end{table}

\begin{table}[t]
\centering
\footnotesize
\caption{Absolute dimensions and main physical parameters of the
components of NSVS0103}
\label{tab:AbsDim}
\begin{tabular}{lcc}
\hline\hline
Parameter & Primary & Secondary\\
\hline
Mass ($M_{\odot}$)  \dotfill & $0.5428 \pm 0.0027$  & $0.4982 \pm 0.0025$\\
Radius ($R_{\odot}$) \dotfill & $0.5260 \pm 0.0028$ & $0.5088 \pm 0.0030$    \\
$\log g$ (cgs)  \dotfill &$ 4.730 \pm 0.005 $ & $4.722 \pm 0.006$ \\
$v_{\rm sync} \sin i$ (km s$^{-1}$)  \dotfill & $72.13\pm 0.38$ & $69.77 \pm 0.41$  \\
$T_{\rm eff}$ (K) \dotfill & 3615 $\pm$ 72 & 3513 $\pm$ 31  \\
$L/L_{\odot}$  \dotfill & 0.0426 $\pm$ 0.0034& 0.0356 $\pm$ 0.0013\\
$M_{\rm bol}$ (mag)  \dotfill & 8.08 $\pm$ 0.25& 8.27 $\pm$ 0.12\\
$M_{V}$ (mag) \dotfill & 9.74 $\pm$ 0.22& 10.04 $\pm$ 0.10\\
\hline\hline
\end{tabular}
\end{table}


\begin{figure}[t]
\epsscale{1.0}
\plotone{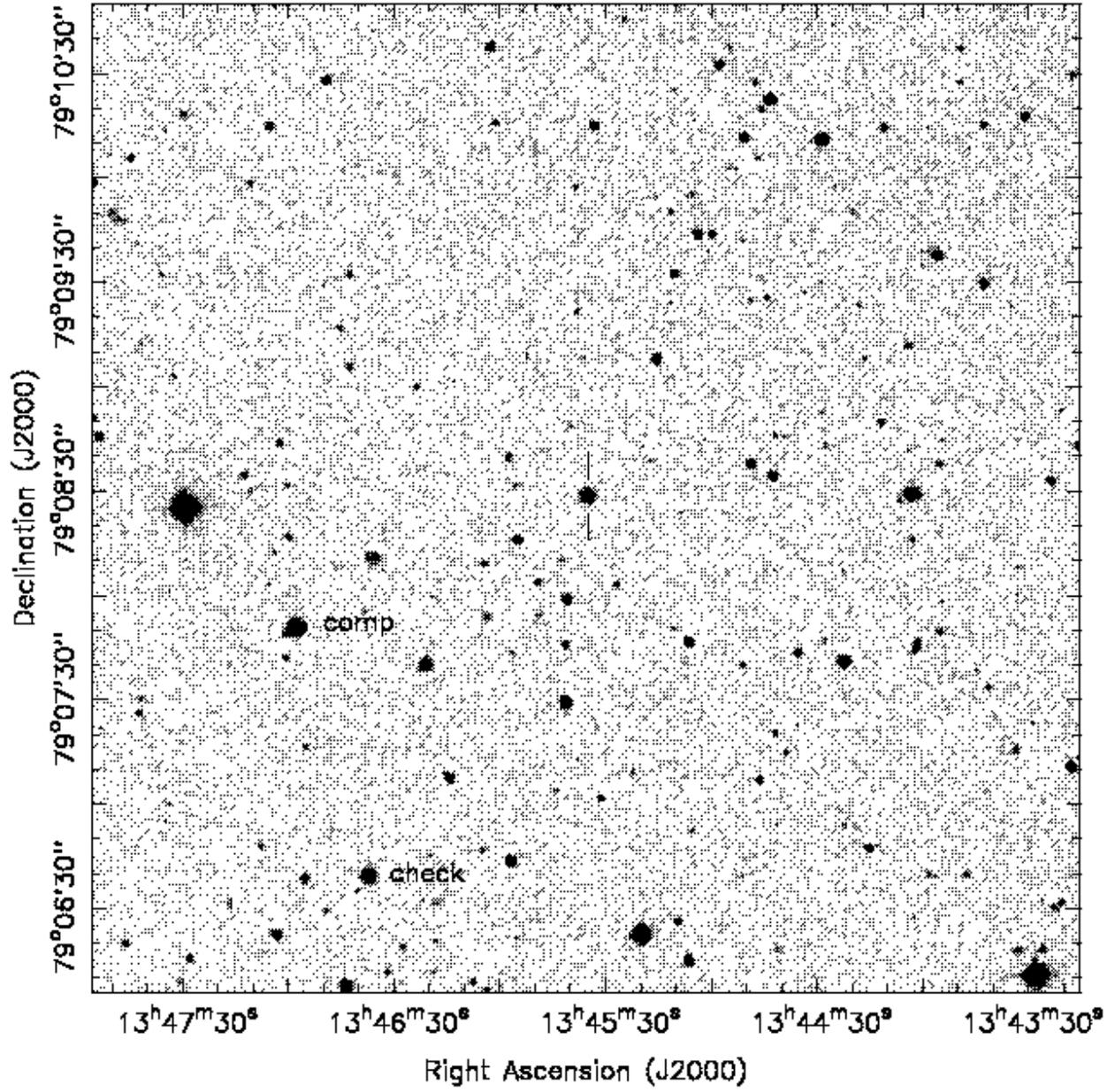}
\caption{Finding chart for NSVS0103, made from the Digitized Sky
Survey.  NSVS0103 is marked with the hash marks.}
\label{fig:fc}
\end{figure}

\begin{figure}[t]
\epsscale{1.0}
\plotone{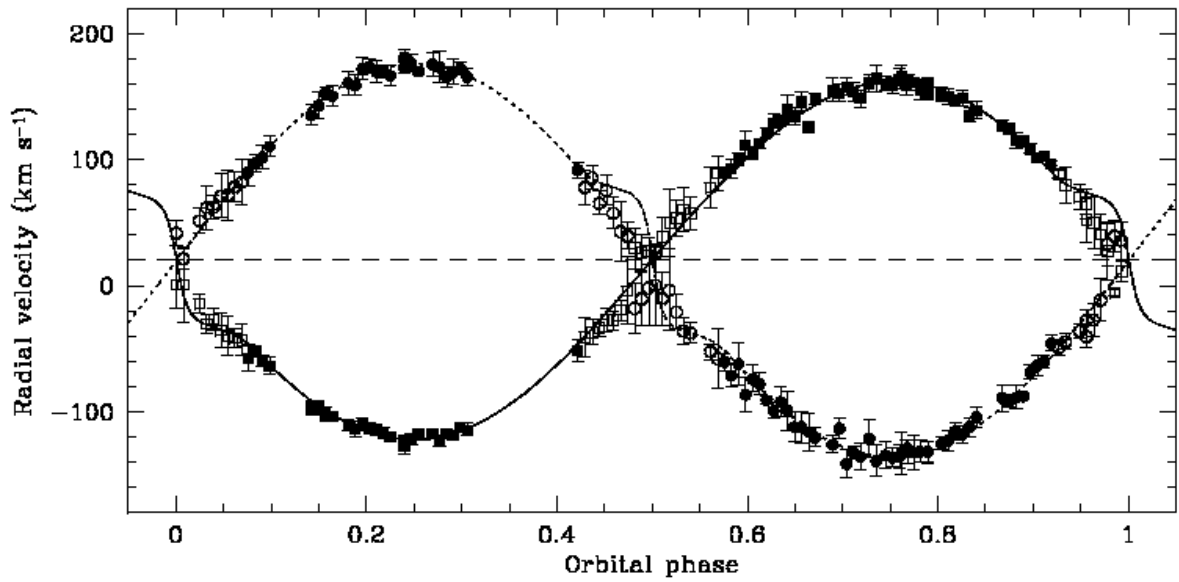}
\caption{Radial velocity curves of NSVS0103. The filled circles and squares correspond, respectively, to the velocities of the primary and the secondary used in the orbital fit in \S 4. The open symbols show points excluded from that fit. The solid and dotted lines represent the orbital solution obtained with ELC.
The dashed line shows the velocity of the center of mass of the system ($\sim$ 19 km/s). These data are available in the electronic edition of this journal (Table 1).}
\label{fig:rv}
\end{figure}

\begin{figure}[t]
\epsscale{1.0}
\plotone{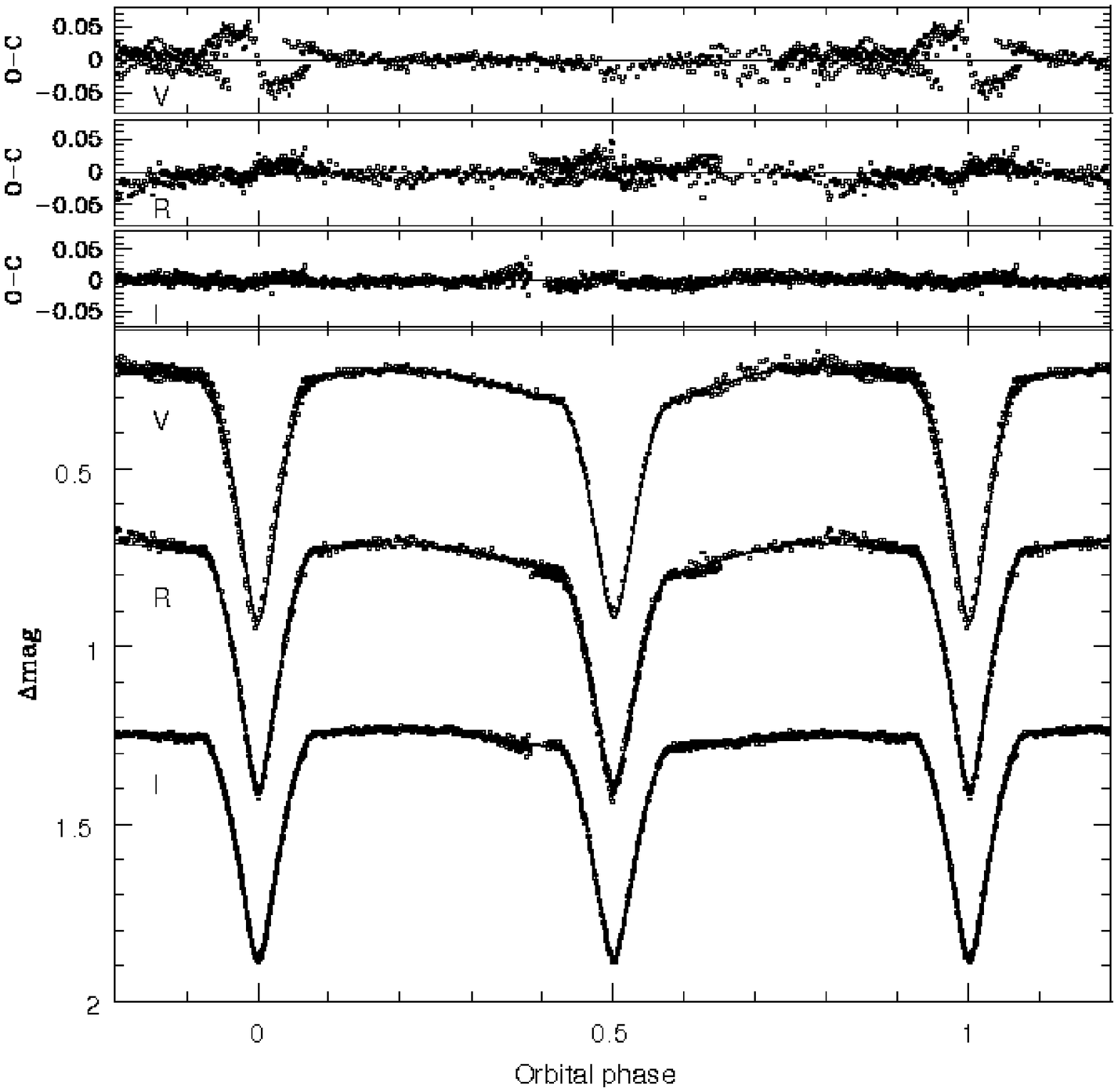}
\caption{V, R, and I light curves from SARA. The dots represent individual observations in each passband. The solid lines show the best fit to the data using ELC. The O-C diagrams on the top show the residuals of those fits. These light curves are available in the electronic edition of this journal (Table 2).}
\label{fig:lcsSARA}
\end{figure}

\begin{figure}[t]
\epsscale{1.0}
\plotone{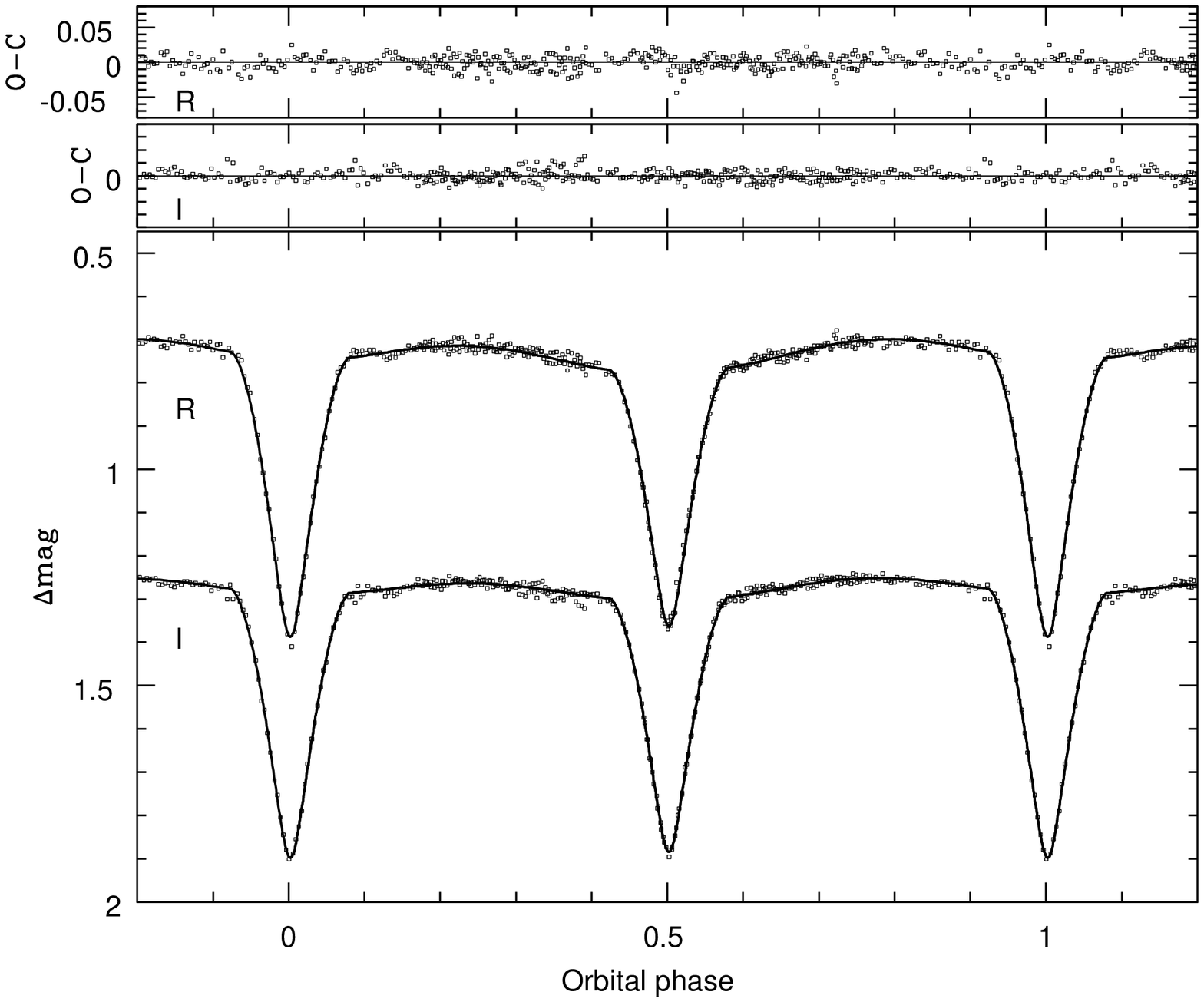}
\caption{R and I light curves from MLO. The dots represent individual observations in each passband. The solid lines show the best fit to the data using ELC. The O-C diagrams on the top show the residuals of those fits. These light curves are available in the electronic edition of this journal (Table 2).}
\label{fig:lcsMLO}
\end{figure}

\begin{figure}[t]
\epsscale{0.9}
\plotone{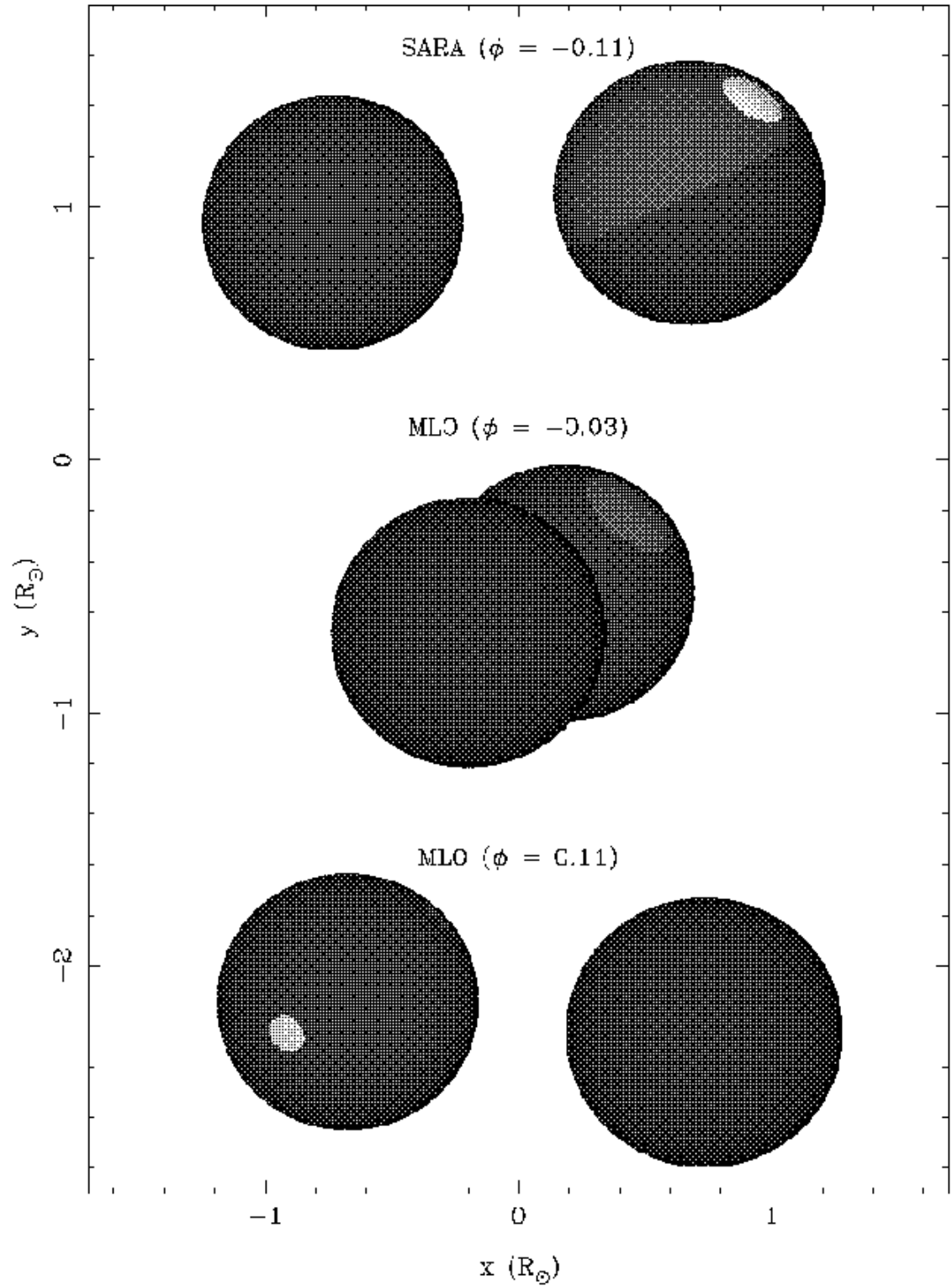}
\caption{Spot models that best reproduce the SARA light curves (top diagram) and the MLO light curves (two bottom diagrams). In both cases the best solution correspond to two bright spots on the primary. The parameters of the spots are listed in Table 4}\label{fig:spot} 
\end{figure}

\begin{figure}[t]
\epsscale{0.9}
\plotone{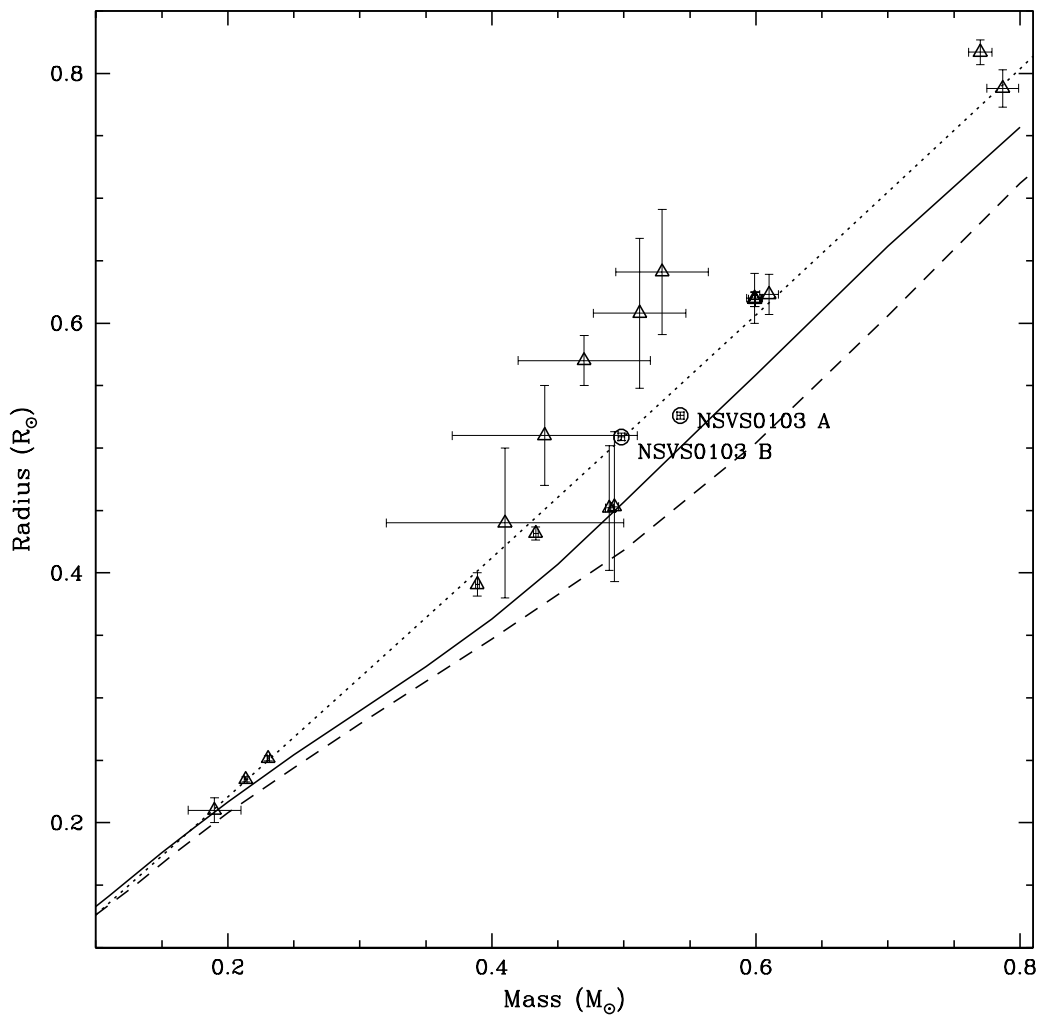}
\caption{Mass-radius relations of stars between 0.1 and 0.8 $M_{\odot}$
predicted by the models of Baraffe et al. (1998) ($solid$ $line$) and Siess 
et al.(2000) ($dashed$ $line$), for an age of 0.35 Gyrs and $Z$ = 0.02, and the empirical relation of Bayless \& Orosz (2006) (($dotted$ $line$). The open triangles show the location of all the low-mass main sequence binaries previously found. The open circles show the two stars in NSVS0103. Each point includes error bars.}\label{fig:mr} 
\end{figure}

\end{document}